\pdfoutput=1 
\documentclass[10pt,a4paper]{article}
\usepackage[utf8]{inputenc}
\usepackage[T1]{fontenc}
\usepackage[english]{babel}
\usepackage{amsmath}
\usepackage{amsfonts}
\usepackage{amssymb}
\usepackage{graphicx}

\usepackage[parfill]{parskip} 
\usepackage{csquotes}
\usepackage{hyperref}
\hypersetup{linktoc=page}
\usepackage{float}
\usepackage{subcaption}
\usepackage[most]{tcolorbox}
\usepackage{booktabs}
\usepackage{enumitem}
\usepackage{tocloft}

\usepackage[backend=biber,style=alphabetic]{biblatex}
\bibliography{quadraticp}

\author{Andrea Barontini\thanks{andrea.barontini@bybaro.it}}
\title{Quadratic Payments with constrained probabilities}
\date{April, 2021}

\begin{document}
	\maketitle
	\begin{figure}[h!]
		\includegraphics[width=\linewidth]{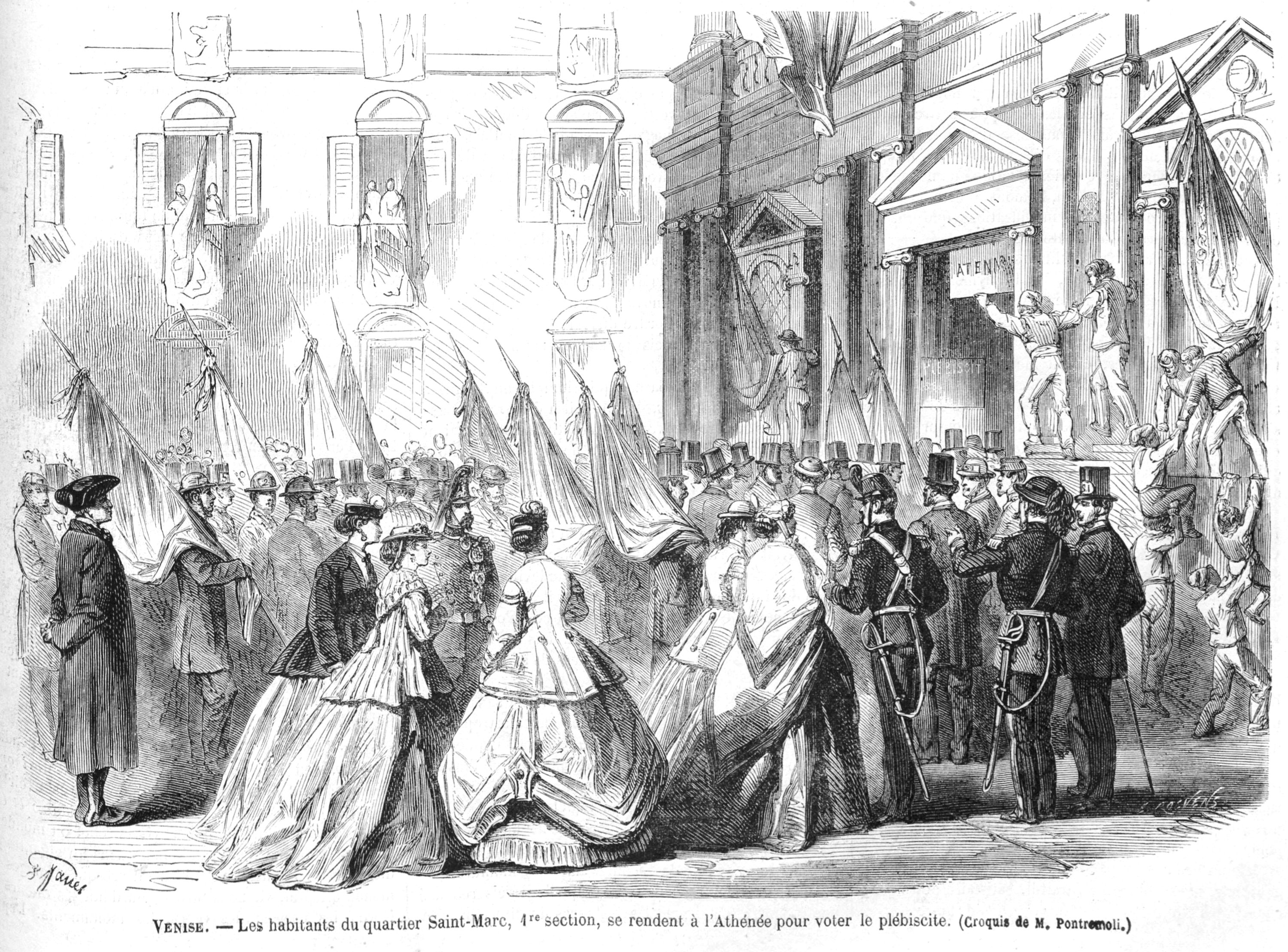}
		\caption{Venice 1866, citizens of San Marco going to vote \cite{Venice}}
		\label{fig:venice}
	\end{figure}
	\begin{abstract}
		Dealing with quadratic payments, marginal probability is usually considered ideally constant, maybe for the sake of initial simplicity. Considering the voting scenario depicted in \cite{Vitalik}, firstly its math foundations are made explicit. Developing a simple referendum model, more realistic outcome probability and marginal probability qualitative shapes are introduced. Enforcing seemingly reasonable assumptions, quadratic payments are then generalized to take into account these new functions shapes, and the way they are still quadratic is discussed. Closing remarks underline the emerging of trade-off constraints not existing in ideal case.
	\end{abstract}

	\newpage
	\tableofcontents
	\listoffigures
	\listoftables
	\printbibliography
	
	\newpage
	\section{Version history}
	This article is the rewrite -to brush up my \LaTeX \ and to use a typesetting environment more suitable for this kind of content- of the one I published on Medium on June 10th, 2020 \cite{Medium}.
	
	Given some new typographic peculiarities, the contents are substantially unchanged apart from a couple of minor additions and some typos corrections:
	\begin{itemize}
		\item subtitle and \enquote{tl;dr} substituted by Abstract;
		\item sentences citing external URLs slightly modified to use bibliographic engine;
		\item added this \enquote{Version history} section;
		\item in \enquote{A generic marginal probability \& generalized quadratic payments} section, inversion notation for $f(i)$ and $c(i)$ is now coherent with the rest of the article;
		\item in equation (7), added  previously missing $_{max}$ subscript;
		\item in last equation before \enquote{And the name?} section, generic $\Delta p(i)$ is now never used in favor of explicit special case $\Delta p$;
		\item \enquote{grey} typos corrected with \enquote{gray};
		\item in the second last equation of \enquote{And the name?} section, changed system members order and restored the correct strict inequality 
	\end{itemize}
	
	\section{An inspiring article}
	Some months ago my geek-attention has been caught by a Vitalik Buterin’s article about quadratic payments and how they could be applied to some everyday choices, e.g. voting \cite{Vitalik}. In many moments of my life I have wondered how much effective democracy and universal suffrage are for ballots on topics whose evaluation could be influenced by specific knowledge and awareness (for example in Italy in 1987 we had a referendum about the use of nuclear energy for civil purposes). So, starting from the article, I have felt the need to elaborate more by myself, to better understand and -why not- try to go deeper.
	
	Before continuing here, I strongly suggest you to take a look at Vitalik’s words if you haven’t done yet, I think they are really inspiring \cite{Vitalik}.
	
	I just quickly recap the concepts about vote pricing, to have a common ground from which going on and to introduce a slightly different notation:
	\begin{itemize}
		\item $i$ is the whole number representing the votes you express, i.e. the votes you have bought;
		\item $c(i)$ is the price, the cost of the $i$-th vote;
		\item $p(i)$ is the probability the referendum will result in your desired outcome when you express i votes (so $p(0)$ is the probability when you don’t take part);
		\item $\Delta p(i)$ is the marginal probability: $\Delta p(i)=p(i)-p(i-1)$, the gain in probability you get when you buy one more vote, the $i$-th, after the already bought $i-1$ votes;
		\item$V$ is how much you value your desired outcome (e.g. if you prefer the \enquote{yes} winning in a referendum, it represents how much important that result is).
	\end{itemize}
	A reasonable condition is that:
	\begin{equation}\label{eq:1}
		\Delta p(i) \cdot V \geq c(i)
	\end{equation}
	stating that the potential value gain you get for $i$-th vote has to be higher or equal to the price of the $i$-th vote: it models the buying threshold of a rational player. Putting it in layman terms:
	\begin{itemize}
		\item $V$ is the value you assign to your desired outcome,
		\item so if the probability of your preferred final result is $1$ then the whole
		referendum values exactly $V$ as well;
		\item if instead the probability of your preferred final result is $0$ then the referendum has no value for you (you could argue that referendum has negative value, but it can be \enquote{absorbed} with an higher $V$ for the case resulting in your desired outcome);
		\item each time you buy a new vote you increase the probability of getting the preferred final result (which, remember, corresponds to the whole referendum valuing $V$) by a factor $\Delta p(i)$, so you don’t want to pay that increase more than the fraction of $V$ (given by $\Delta p(i)V$) that you have potentially gained.
	\end{itemize}
	Of course the correspondence between actual (not the perceived) value and probability lies in frequency interpretation of the latter and return of investment cannot be guaranteed: a referendum happens only once, you really don’t have many occurrences of it determining \enquote{success cases} over \enquote{total cases}, and outcome is always binary: \enquote{yes} wins or \enquote{no} wins, no fuzziness there! That’s why I have used the \enquote{potential} and \enquote{potentially} words earlier. Nevertheless the goal here is to build incentives to make influence of each stakeholder proportional to their perceived value V, and previous mathematical stuff works for that.
	
	I also want to underline a couple of Vitalik’s assumptions, I guess for the sake of simplicity given the introductory nature of his article:
	\begin{itemize}
		\item $\Delta p(i)$ is always considered constant: $\Delta p(i) \equiv \Delta p$\\
		You can suspect that things get more complicated outside of ideal case when he writes: \enquote{\textit{[...] though eventually the gains will decrease as the probability approaches 100\% [...]}} , but it seems to be the only reference about it;
		\item \textit{influence} isn’t actually defined to keep the scope of the reasoning wide, however we can observe that with constant $\Delta p$ we are in a special case where the \textit{influence} is simply the number of votes $i$ that you have bought.
	\end{itemize}
	That said, the constraint \ref{eq:1} becomes:
	\begin{equation}\label{eq:2}
		\Delta p \cdot V \geq c(i)
	\end{equation}
	The goal is to enforce \ref{eq:2} to act as incentive to get an \textit{influence} proportional to $V$, so in our special case we’ll want the maximum number of bought votes $i_{max}$ to be proportional to $V$.
	
	Different types of cost function $c(i)$ are considered; First case:
	\begin{equation*}
		c(i) \equiv c \quad \Longrightarrow \quad \Delta p \cdot V \geq c \quad \Longrightarrow \quad V \geq \frac{c}{\Delta p}
	\end{equation*}
	so when $V$ reaches $c/\Delta p$ every vote we want to buy is worth its price: this means $i_{max}$ is unlimited if you can afford $c$ price for each of them: everyone who has a lot of funds to spend has a lot of influence, regardless of how much the desired outcome is valued ($V$).
	
	Second case:
	\begin{equation*}
		c(i)=
		\begin{cases}
			c & i = 1\\
			\infty & i > 1
		\end{cases}
	\end{equation*}
	After the first vote, all the others have infinite cost, meaning you can buy just one, again regardless of $V$.
	\begin{figure}
		\includegraphics[width=\linewidth]{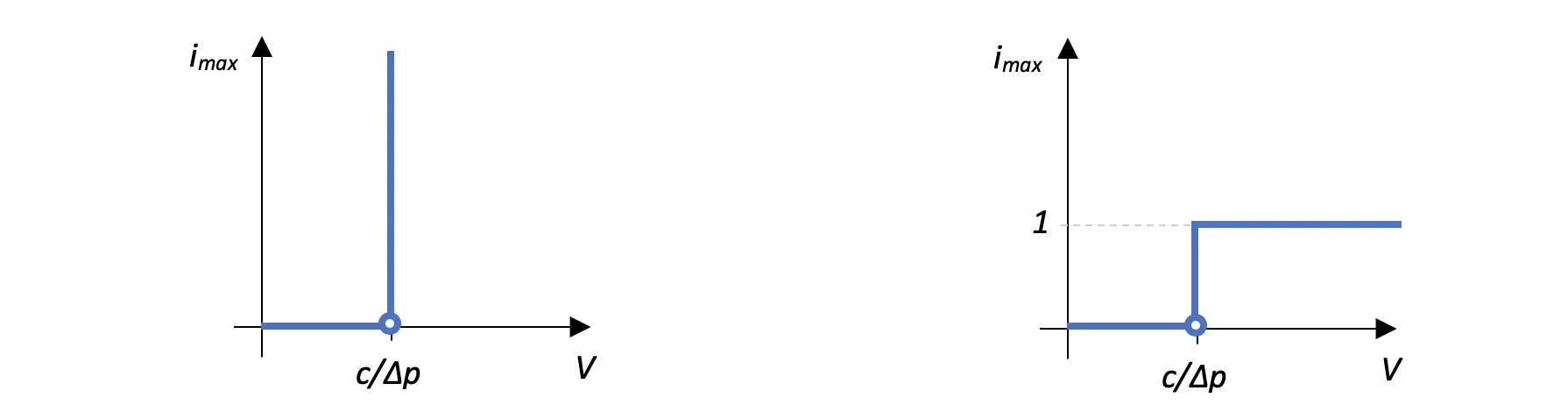}
		\caption{Plutocratic \& Democratic Cost Functions}
		\label{fig:prequadratic}
	\end{figure}

	We could say first case is too plutocratic, the second one too democratic!
	
	That’s where quadratic payments magic comes on stage! If we use a cost
	function linear in $i$ (Vitalik uses a smart heuristic reasoning to derive it):
	\begin{equation}\label{eq:3}
		c(i) = c \cdot i
	\end{equation}	
	from the constraint \ref{eq:2} we get:
	\begin{equation*}
		\Delta p \cdot V \geq c \cdot i \quad \Longrightarrow \quad i \leq \Delta p \frac{V}{c} \quad \Longrightarrow \quad i_{max} = \left \lfloor \Delta p \frac{V}{c}\right \rfloor
	\end{equation*}
	\begin{figure}[H]
		\includegraphics[width=\linewidth]{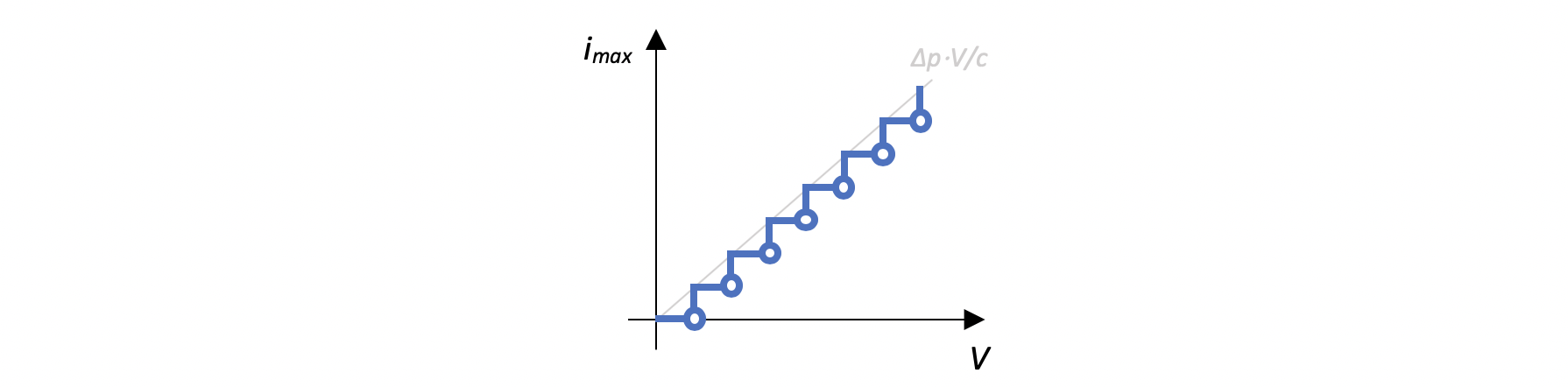}
		\caption{Linear Cost Function}
		\label{fig:quadratic}
	\end{figure}
	the stairs-like profile derives from $i_{max}$ needing to be a whole number, but linearity in whole $\Delta p \cdot V/c$ is plain! And if we calculate how much we spend to buy $i_{max}$ votes:
	\begin{equation*}
		c(1)+c(2)+\dots+c(i_{max})=c(1+2+\dots+i_{max})=c \frac{i_{max}(i_{max}+1)}{2} \in \mathcal{O}(i_{max}^2)
	\end{equation*}
	which is why we call them \enquote{quadratic payments}.
	
	More or less, very summarized and with a little of more math here and there, this is the core of what you can find about voting in Vitalik’s article... now let’s try to make a few steps ahead.
	
	\section{Let’s play with math!}
	Let’s try to derive in a formal way the linear cost function \ref{eq:3}. From \ref{eq:2} we know:
	\begin{equation*}
		c(i) \leq \Delta p \cdot V
	\end{equation*}
	let’s make two assumptions (that of course will have to be verified and will anyway limit the generality of the result):
	\begin{itemize}
		\item $c(i)$ is an invertible function
		\item it's inverse $c^{-1}(i)$ is monotonically increasing (to avoid problems with inequality sign)
	\end{itemize}
	which, by the way, can be condensed requiring $c(i)$ to be monotonically strictly increasing; so we can write:
	\begin{equation*}
		c^{-1}(c(i)) \leq c^{-1}(\Delta p\cdot V) \quad \Longrightarrow \quad i \leq c^{-1}(\Delta p\cdot V)
	\end{equation*}
	we have a condition limiting $i$, what we wanted. Then we require the $i$-upper bound to be proportional to $V$ (remember that we are seeking a $c(i)$ which makes $i_{max} \propto V$):
	\begin{equation*}
		i \leq c^{-1}(\Delta p\cdot V) = K \cdot V \quad \textrm{,} \quad K>0
	\end{equation*}
	from simple algebra (multiplying and dividing by $\Delta p$) it follows that:
	\begin{equation*}
		c^{-1}(\Delta p\cdot V) = \left(\frac{K}{\Delta p}\right) (\Delta p\cdot V)
	\end{equation*}	
	so we can define:
	\begin{equation*}
		c^{-1}(\bullet) \triangleq \frac{K}{\Delta p}\bullet
	\end{equation*}		
	we invert and get:
	\begin{equation}\label{eq:4}
		c(i)=\frac{\Delta p}{K}i
	\end{equation}
	Last step, we note \ref{eq:4} is monotonically strictly increasing, so it’s an acceptable result because it respects the previous assumptions. And, as expected, it confirms \ref{eq:3} setting $K= \Delta p/c$. So we have:
	\begin{equation}\label{eq:5}	
		\begin{cases}
			i \leq K \cdot V \quad \Longrightarrow \quad i_{max} = \lfloor KV \rfloor\\
			c(i) = \displaystyle\frac{\Delta p}{K}i 
		\end{cases}
	\end{equation}
	It’s also useful to underline a few technicalities of the formal derivation:
	\begin{itemize}
		\item $K$, by which we have imposed the proportionality between influence and perceived value, also acts as a result’s degree of freedom defining both the maximum number of bought votes and the cost of the first one (the cheaper)... and it seems to be an unbounded degree of freedom (apart from, of course, being positive).
		\item The derivation cannot say anything about not monotonically increasing cost functions because only if it’s monotonically increasing we can, inverting, transform condition on cost $c(i)$ into condition on vote index $i$; however\dots
		\item \dots it’s not a huge limit by itself because we have an ab-initio more serious lack of generality given by $\Delta p(i) \triangleq \Delta p$ (if not, we should invert ${f(i) \triangleq c(i) / \Delta p(i)}$ and we couldn’t obtain a strictly defining condition for $c(i)$)
	\end{itemize}
	By now, it seems obvious we have to dig into the \enquote{shape} of $\Delta p(i)$ in \enquote{real life} to make any educated guess on how to proceed.
	
	\section{A simple referendum model}
	So I brushed up my old interest in voting effectiveness and I tried to come up with a simple ballot model.\\
	Let’s imagine that we are close to a referendum (a ballot with only two possible outcomes: \enquote{yes} or \enquote{no}): supporting the \enquote{yes}, we have done a statistical research on voters and we have got a voting prediction for the \enquote{average voter}. What I mean is that, instead of dealing with many different voters, each one with a different probability to vote \enquote{yes}, in our calculations we will use the average voter and the voting prediction associated to him (ok, it’s a rough model, but we have to start from somewhere). We have:
	\begin{itemize}
		\item $n$: number of voters
		\item $y$: probability the average voter will vote \enquote{yes} (our voting prediction)
		\item $p(y,n)$: probability the referendum outcome will be \enquote{yes}
	\end{itemize}
	resulting in:
	\begin{equation*}
		p(y,n)=\sum_{d=\left \lfloor n/2 \right \rfloor +1}^{n} \binom{n}{d} y^{d} (1-y)^{n-d}
	\end{equation*}
	Let me try to convince you this is a reasonable model.\\
	Each addend of the summation takes into account cases in which \enquote{yes} supporters are more than \enquote{no} supporters, starting from minimal difference (1 or 2 votes, depending if $n$ is odd or even) and ending with all voters choosing the \enquote{yes}.\\
	In each iteration $d$ is the number of \enquote{yes} voters, the remaining $n-d$ the number of \enquote{no} voters, and the binomial coefficient returns the number of different \enquote{order} combinations of \enquote{yes} and \enquote{no} votes (e.g. $yynyn \dots$, $yyynn\dots$, $nynyy\dots$, $\dots$).\\
	Plotting this function gives some by-itself interesting insights into the model:
	\begin{figure}[H]
		\centering
		\includegraphics[height=0.33\textheight]{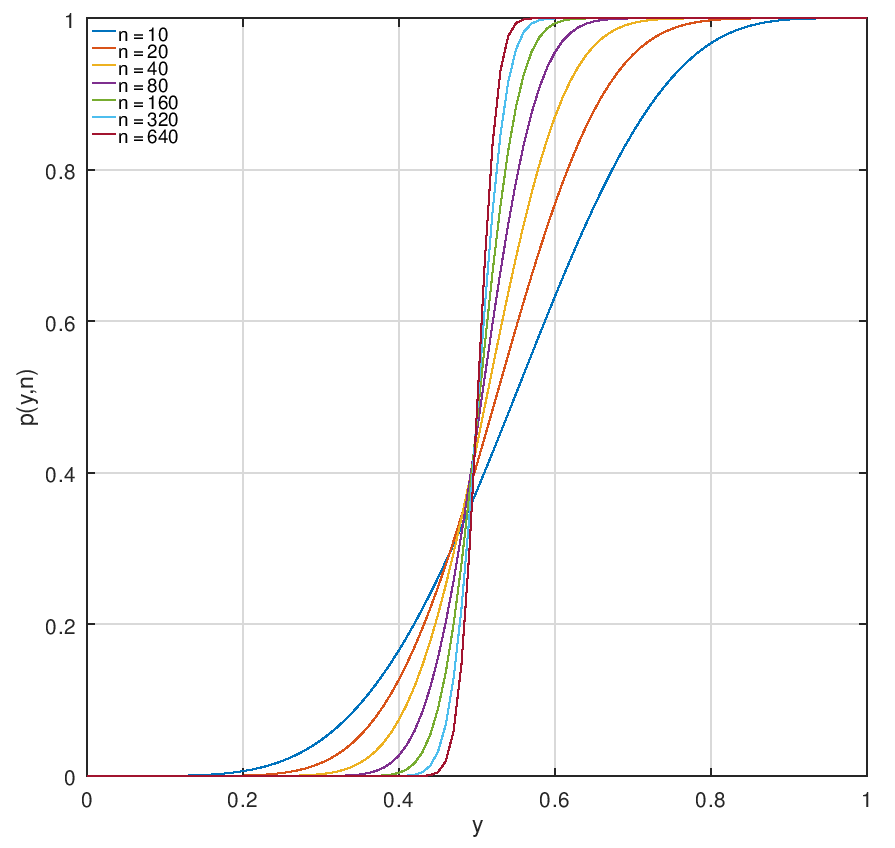}
		\caption{Simplest referendum model}
		\label{fig:simplest}
	\end{figure}	
	We can see that when the number of voters $n$ grows, the curve tends to a step function polarizing the outcome (\enquote{yes} or \enquote{no}) on the sides of a discontinuity of $p(y,n)$ at $y=0.5$: when there are a lot of voters, also a very small bias in vote preference will cause the referendum result to be \enquote{yes} or \enquote{no} for certain, depending on the direction of the bias. Luckily, it’s what we expect from a referendum: even if there’s a lot of uncertainty, we want one party to win even if by few votes.
	
	So let’s look what happens for exactly $y=0.5$ (and a couple of other values) when $n$ grows:
	\begin{figure}[H]	
		\centering
		\includegraphics[height=0.33\textheight]{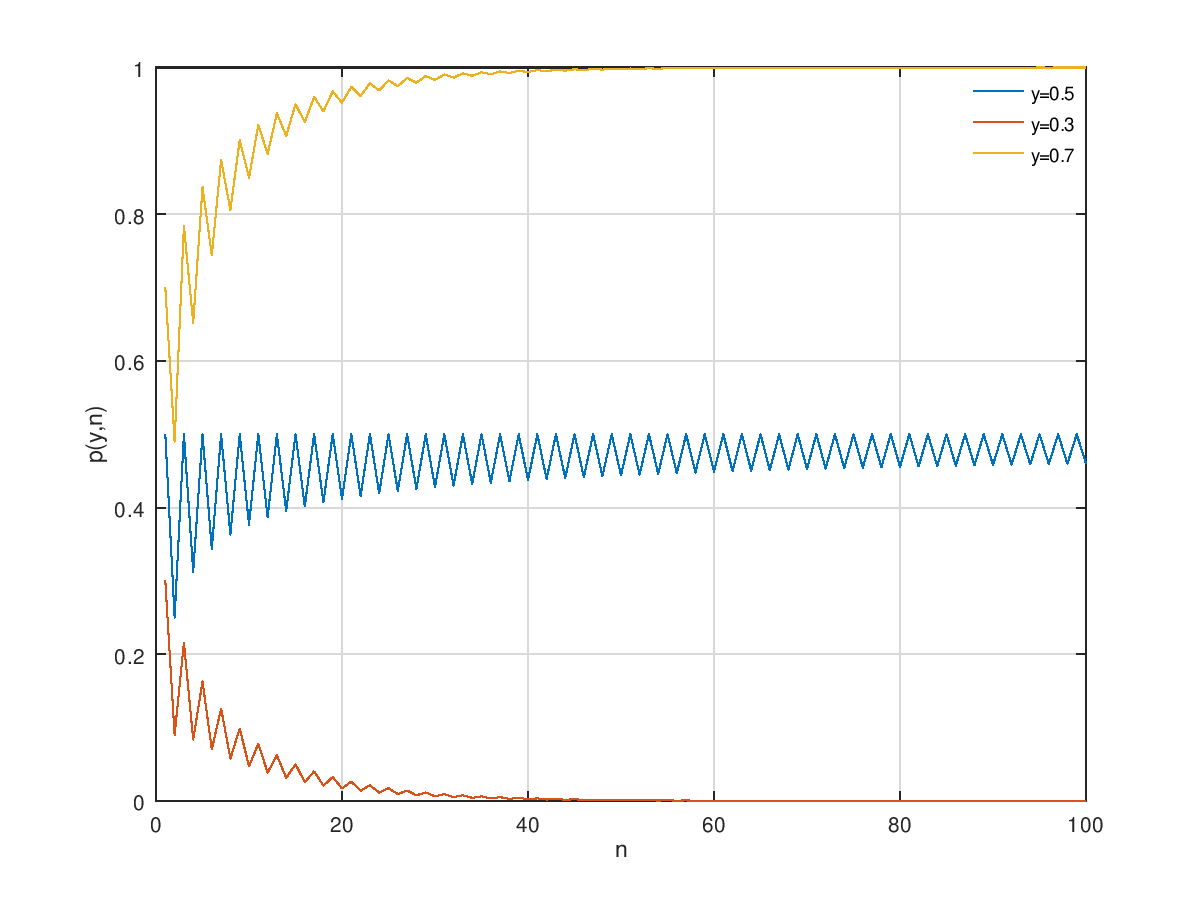}
		\caption{Growing number of voters}
		\label{fig:growing}
	\end{figure}
	
	As we have seen earlier, for $y \neq 0.5$, $p(y,n)$ quickly tends to 1 or 0; for $y = 0.5$ it seems converging to 0.5 but with an at-first-unexpected sawtooth-like profile (visible in the leftmost parts of the other two curves as well): what is that? Just a tip: think to roulette-gambling$\dots$\\
	\dots\\
	Ok, time’s up: it depends on $n$ being alternatively odd and even: when even, the referendum final outcome could be \enquote{yes}, \enquote{no}... but a break-even could also be possible, and of course votes combinations leading to it will lower the \enquote{yes} probability, for every $y$ (like zero in roulette makes betting on a color a less than 50\% affair).
	
	Good, our model seems reasonable: so, remembering our purpose of exploring a more \enquote{real} (or at least a \enquote{less unreal}) $\Delta p(i)$, let’s take a step further introducing vote buying. Our function becomes:
	\begin{equation*}
		p(y,n,i)=
		\begin{cases} 
			\displaystyle
			\sum_{d=\left \lfloor n/2 \right \rfloor +1-i}^{n-i} \binom{n-i}{d} y^{d} (1-y)^{n-i-d} & \quad i \leq \left \lfloor n/2 \right \rfloor +1\\
			\qquad \qquad \qquad \qquad 1 & \quad i > \left \lfloor n/2 \right \rfloor +1
		\end{cases}
	\end{equation*}
	where, as earlier, $i$ is the number of votes bought for \enquote{yes}. The new expression reflects that:
	\begin{itemize}
		\item the total number of votes whose probabilities need to be taken into account is not $n$ anymore, but $n-i$ in fact (each bought vote has probability 1 so no need to appear in calculus);
		\item if \enquote{yes} outcome has an \enquote{$i$ votes initial treasure}, we also need to sum the probabilities of cases which lack -compared to previous expression- up to $i$ \enquote{yes} votes (that’s why the change in summation lower bound);
		\item the inequality condition on $i$ is the mathematical way to guarantee all quantities involving it are positive or zero, but also the formalization of the obvious fact that it’s enough to buy half of the votes plus one to be sure of \enquote{yes} winning.
	\end{itemize}
	Fixing $n=100$ and plotting between $0 \leq i \leq 60$:
	\begin{figure}[H]	
		\centering
		\includegraphics[height=0.26\textheight]{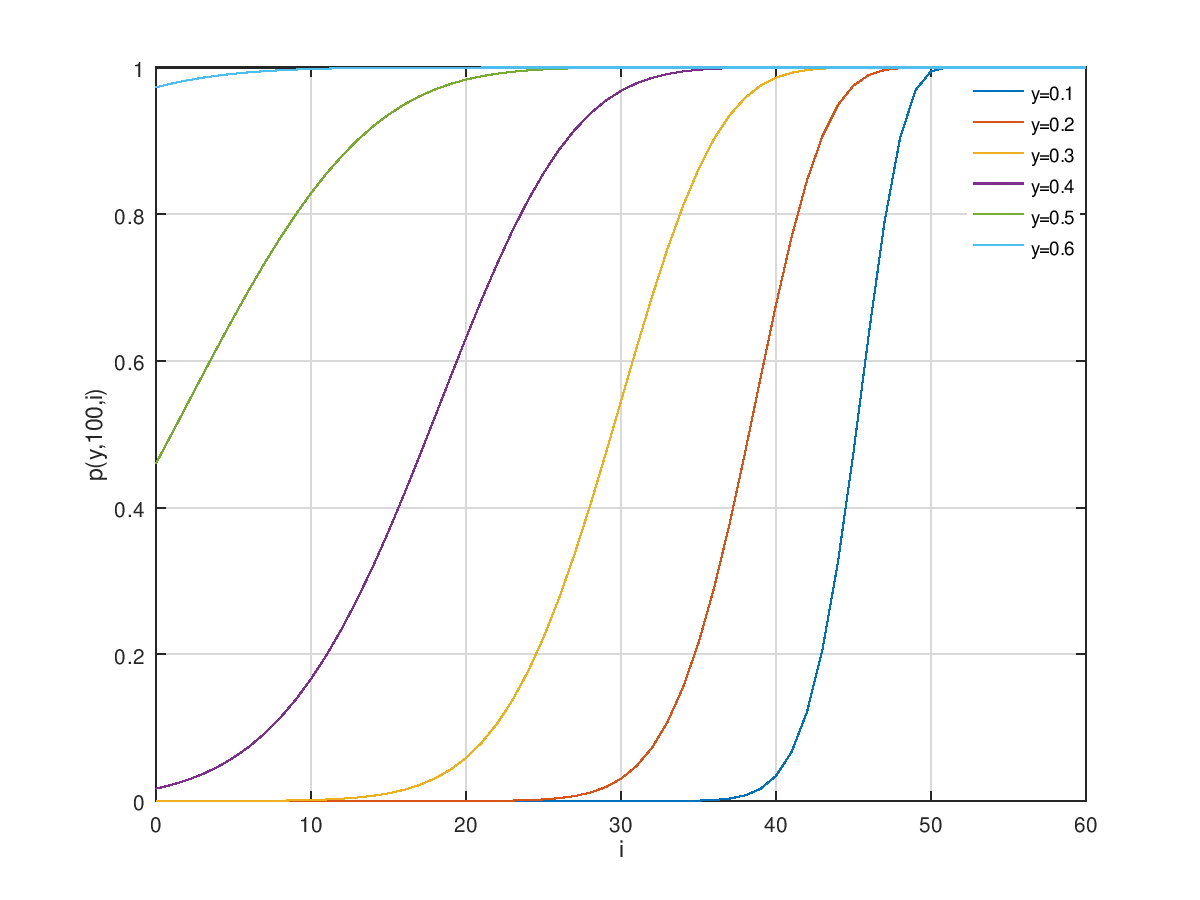}
		\caption{Referendum model with votes-buying}
		\label{fig:buying}
	\end{figure}
	Of course higher $y$ (probability the average voter will choose \enquote{yes}), sooner $p(y,100,i)$ reaches 1 while $i$ grows; for $y > 0.6$ buying votes is almost useless (remember discontinuity around $y = 0.5$ for high number of voters), and by $i = 51$ ($= n/2 +1$) all curves have reached highest probability.\\
	So if a very uncertain referendum involves 100 voters and our statistical research can only establish that their penchant for \enquote{yes} falls between 40\% and 50\%, then graph above tells us we have to buy at least 9 votes (and no more than 23 needed) to have an higher than 80\% probability of \enquote{yes} winning (just check for which i the curves for $y=0.4$ and $y=0.5$ reach the 0.8 height).
	
	By the way, it’s obvious that $\Delta p(i) = p(i)-p(i-1)$ cannot be constant, but let’s plot it to see its shape:
	\begin{figure}[H]	
		\centering
		\includegraphics[height=0.33\textheight]{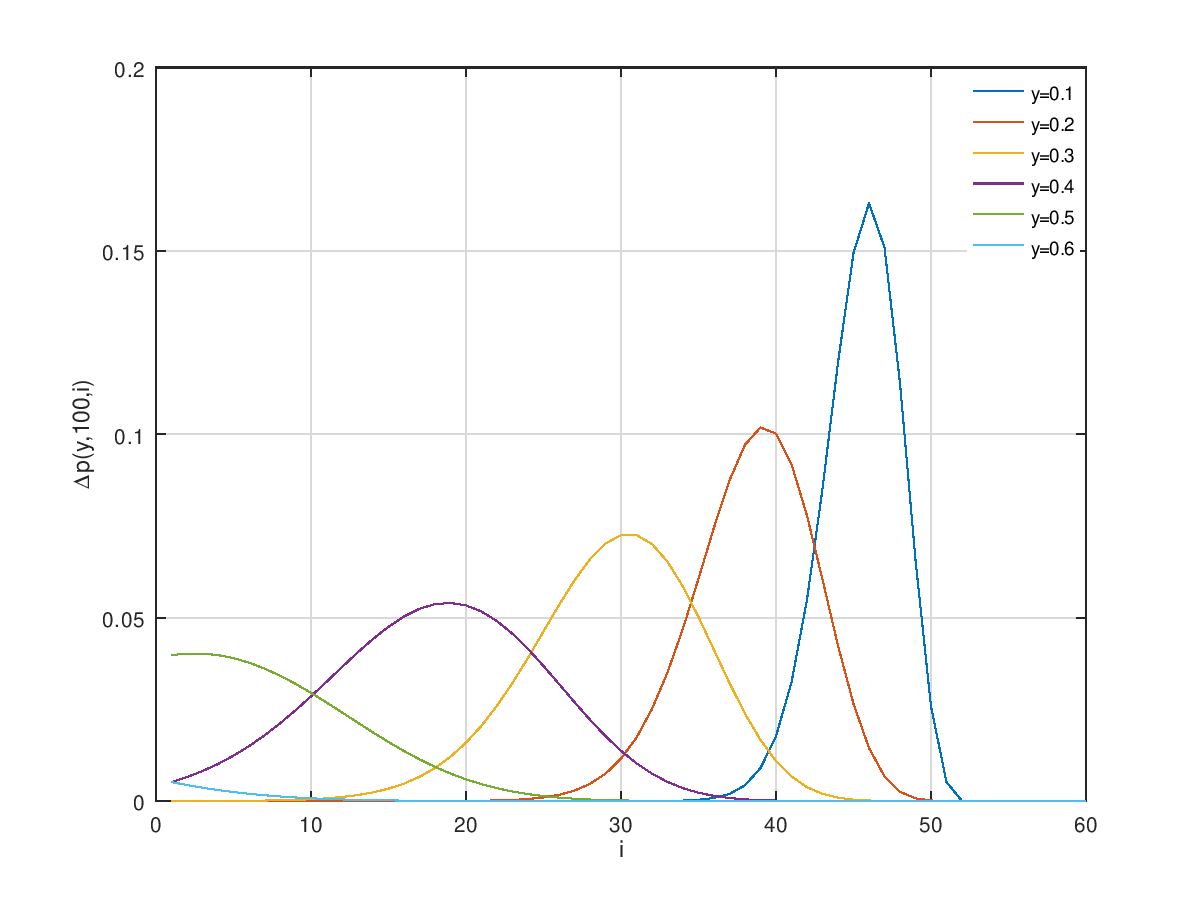}
		\caption{\enquote{Real} marginal probability}
		\label{fig:marginalp}
	\end{figure}
	Definitely not constant!\\
	We could play a lot this way: if you want on my GitHub there’s a repository \cite{Github} where I have uploaded some unleashed Octave code I have used to generate the above graphs and a quite big archive (almost 1GB overall) of precomputed values of $p(y,n,i)$ for the $101 \times 1000 \times 502$ lattice domain defined by:
	\begin{equation*}
		\begin{cases} 
			y \in(0, 0.01, 0.02, \dots, 1)\\
			n \in(1, 2, 3, \dots, 1000)\\
			i \in(0, 1, 2, \dots, 501)
		\end{cases}
	\end{equation*}
	Feel free to play with it, if you want. But here it’s time to extrapolate a few general properties of $p(i)$ and $\Delta p(i)$ (inspired by, but independent from, our referendum model).
	
	\section{A generic marginal probability \& generalized quadratic payments}
	Let’s recap what we have discovered about our probabilities:
	\begin{itemize}
		\item $p(i)$ is monotonically strictly increasing between 0 and 1;
		\item $\Delta p(i)$ is consequently positive but in general neither constant nor monotonically increasing (that’s all we know about our generic marginal probability).
	\end{itemize}
	It seems now we haven’t a lot to try a formal derivation as we did earlier when marginal probability $\Delta p(i) \equiv \Delta p$, because (check \enquote{technicalities} after \ref{eq:5}):
	\begin{itemize}
		\item $\Delta p(i)$ isn’t constant in $i$ so we should invert $f(i) \triangleq c(i)/\Delta p(i)$ \dots
		\item ...but, wanting to deal with a general case, then we haven’t enough constraints on $\Delta p(i)$ to be able to derive (please note that here $[\quad]^{-1}$ is the inversion, not a power):
		\begin{equation*}
			f^{-1}(i) = \left[ \frac{c(i)}{\Delta p(i)}\right]^{-1} \longrightarrow \quad c^{-1}(i) \quad \longrightarrow \quad c(i)
		\end{equation*}
	\end{itemize}
	We have to try to proceed in a wily way. Let’s begin trying to find an expression for $i_{max}$.
	
	When marginal probability was constant, degree of influence was simply given by the number of bought votes $i$, so making influence proportional to perceived value $V$ led to:
	\begin{equation*}
		i \leq KV \quad \Longrightarrow \quad i_{max} = \lfloor KV \rfloor
	\end{equation*}
	Now each bought vote brings an influence increase $\Delta p(i)$, so to impose influence-value proportionality we should write:
	\begin{equation}\label{eq:6}
		\sum_{i=1}^{i_{max}} \Delta p(i) = K_{2}V \quad \textrm{,} \quad K_{2}>0
	\end{equation}
	The above expression isn’t rigorous as-is, other constraints will apply apart from $K_{2}$ positiveness, we will deal with them in a few lines.\\
	By the way, now we use $K_{2}$ because when earlier we introduced $K$ we omitted $\Delta p$ from influence evaluation; however adding just one more line to that formal derivation we could have written instead:
	\begin{equation}\label{eq:7}
		\Delta p \cdot i \leq K_{2}V \quad \Longrightarrow \quad i_{max} = \left\lfloor \frac{K_{2}}{\Delta p} V \right\rfloor
	\end{equation}
	From which it follows (it will be useful later) that $K = K_{2} / \Delta p$.
	
	Returning to our summation, to expand it we note that:
	\begin{equation*}
		\sum_{i=1}^{m} \Delta p(i) = \big(p(1)-p(0)\big) + \big(p(2)-p(1)\big) + \dots + \big(p(m)-p(m-1)\big) = p(m)-p(0)
	\end{equation*}
	so:
	\begin{equation*}
		\sum_{i=1}^{i_{max}} \Delta p(i) = K_{2}V \quad \Longrightarrow \quad p(i_{max})-p(0) =  K_{2}V \quad \Longrightarrow \quad p(i_{max}) = K_{2}V + p(0)
	\end{equation*}
	Having a probability on the left side of the equation limits the permitted values of right side: here it is one more constraint on $K_{2}$ (differently from $K$ which was unbounded instead):
	\begin{equation*}
		K_{2}V+p(0)<1 \quad \Longrightarrow \quad K_{2} < \frac{1-p(0)}{V}
	\end{equation*}
	it’s a strict inequality (equality not allowed!) because we will want to invert $p(i)$ and, as we have discovered with our referendum model, probability function saturates when it reaches 1 (remember, to be sure of \enquote{yes} winning buying $n/2 + 1$ votes was enough): so invertibility of $p(i)$ is possible only in $[0,1[$.\\
	So, recapping, we got:
	\begin{equation*}
		p(i_{max}) = K_{2}V + p(0) \quad \textrm{,} \quad 0 < K_{2} < \frac{1-p(0)}{V}
	\end{equation*}
	However we need another constraint because $i$ and $i_{max}$ are whole numbers and:
	\begin{itemize}
		\item $p^{-1}(p(i_{max})) \equiv i_{max}$ for sure \dots
		\item \dots but $p^{-1}(K_{2}V+p(0))$ isn’t necessarily a whole number
	\end{itemize}
	Unluckily we cannot introduce a further explicit constraint on $K_{2}$ because we don’t actually know $p^{-1}(\quad)$, so we cannot calculate which values for $K_{2}$ would make the right side of the equality a whole number. However, instead of searching which values of $K_{2}$ should be discarded, we can choose to accept all of them and to lead the outcomes of this \enquote{permissiveness} back to the allowed values. How? With the floor function:
	\begin{equation*}
		i_{max} = p^{-1}(p(i_{max})) = \left\lfloor p^{-1}(K_{2}V+p(0)) \right\rfloor \quad \textrm{,} \quad 0 < K_{2} < \frac{1-p(0)}{V}
	\end{equation*}
	Needing to insert (in an order-preserving way) each real number $p^{-1}(K_{2}V + p(0))$ into one of a set of equivalence classes labeled by whole numbers, each real value has two main whole numbers it can be mapped to: the greatest less than itself (given by the floor function) and the least greater than itself (given by the ceiling function); why have we chosen the first one? Because we are seeking the maximum whole $i$, so the real number we get is a sort of upper-bound, a value which cannot be exceeded by the whole number we need.\\
	If not yet convinced, let’s check if this $i_{max}$ expression falls back to \ref{eq:7} when ${\Delta p(i) \equiv \Delta p}$:
	\begin{equation*}
		p(i): \rho = \Delta p \cdot i + P \quad \Longrightarrow \quad p^{-1}(\rho): i = \frac{\rho - P}{\Delta p}
 	\end{equation*}
 	so:
 	\begin{equation*}
 		i_{max}=\left\lfloor p^{-1}(K_{2}V+p(0)) \right\rfloor = \left\lfloor \frac{(K_{2}V+P)-P}{\Delta p} \right\rfloor = \left\lfloor \frac{K_{2}V}{\Delta p} \right\rfloor = \left\lfloor KV \right\rfloor
	\end{equation*}
	which confirms that floor function is the right choice, allowing particular $i_{max}$ expression to be derived from the general one.\\
	By the way, you have perhaps noted that the way we have inverted $p(i)$ implies both $p^{-1}(i)$ and $p(i)$ being $\mathbb{R} \rightarrow \mathbb{R}$ functions; no problem even if $p(i)$ was initially defined as $\mathbb{N} \rightarrow \mathbb{R}$ function, because we can always extend it to a $\mathbb{R} \rightarrow \mathbb{R}$ one (with a polyline if nothing better is suitable, we don’t need to calculate derivatives).
	
	Now that we know how to calculate $i_{max}$ for a generic marginal probability, it’s time to focus on cost function $c(i)$.\\
	Because of fairness considerations it would be sound for $c(i)$, once we consider a specific value $i$, to be proportional to $\Delta p(i)$: a vote price should be linear in probability gain it allows, given all other conditions. This means the structure of cost function should be: $c(i)= \Delta p(i) g(i)$.\\
	Cost function enforces $i_{max}$ limit: it must be unfavorable to buy more than $i_{max}$ votes, so we want $g(i)$ to keep $c(i)$ increasing with $i$, until we get:
	\begin{equation*}
		c(i)=\Delta p(i) \cdot g(i) > \Delta p(i) \cdot V \quad \textrm{,} \quad i > i_{max}
	\end{equation*}
	Note we are just saying that we want to satisfy -even if in its dual reformulation- condition \ref{eq:1}. So:
	\begin{equation*}
		c(i)=\Delta p(i) \cdot g(i) \leq \Delta p(i) \cdot V \quad \Longrightarrow \quad g(i) \leq V \quad \textrm{,} \quad i \leq i_{max}
	\end{equation*}
	Applying $g^{-1}(\quad)$ to both sides we obtain:
	\begin{equation*}
		i \leq g^{-1}(V) \quad \textrm{,} \quad i \leq i_{max}
	\end{equation*}
	(inversion and inequality sign are ok because sought $g(i)$ will be an increasing function with positive domain and codomain). We now have two inequalities with $i$ on the left side, comparing them:
	\begin{equation*}
		\left\lfloor g^{-1}(V) \right\rfloor = i_{max}
	\end{equation*}
	with floor function appearing, as usual, to handle $ i_{max}$ whole-ness.\\ Remembering  $i_{max}$ formula, we get an explicit expression for $g^{-1}(V)$:
	\begin{equation*}
		g^{-1}(V): i = p^{-1}(K_{2}V+p(0))
	\end{equation*}
	Inverting it:
	\begin{equation*}
		g(i)=\frac{p(i)-p(0)}{K_{2}}	
	\end{equation*}	
	strictly increasing, with positive domain and codomain. Great!\\ 
	We have discovered that:
	\begin{equation*}
		c(i)=\Delta p(i) \frac{p(i)-p(0)}{K_{2}}	
	\end{equation*}	
	Let’s double-check that \ref{eq:1} is satisfied (\enquote{double} because it should, having started the derivation imposing it):
	\begin{equation*}
		c(i)=\Delta p(i) \frac{p(i)-p(0)}{K_{2}} \leq \Delta p(i) \cdot V \quad \Longrightarrow \quad \frac{p(i)-p(0)}{K_{2}} \leq V
	\end{equation*}
	So, as expected:
	\begin{equation*}
		p(i) \leq K_{2}V+p(0) \quad \Longrightarrow \quad i \leq \left\lfloor p^{-1}(K_{2}V+p(0)) \right\rfloor = i_{max}
	\end{equation*}
	Good, and now last step, evaluating $c(i)$ when $\Delta p(i) \equiv \Delta p$:
	\begin{align*}
			p(i) &= \Delta p \cdot i + P\\
			c(i) &= \Delta p \frac{p(i)-p(0)}{K_{2}}=\Delta p \frac{(\Delta p \cdot i + P)-P}{K_{2}} = \frac{\Delta p^{2}}{K_{2}}i = \frac{\Delta p}{K}i
	\end{align*}
	Success!

	\section{And the name?}
	In the beginning, calculating how much we spend to buy $i_{max}$ votes when
	$\Delta p(i) \equiv \Delta p$, we got:
	\begin{equation*}
		\sum_{i=1}^{i_{max}} c(i) = c \frac{i_{max}(i_{max}+1)}{2} \in \mathcal{O}(i_{max}^2)
	\end{equation*}
	Which, considering \ref{eq:4}, for the sake of thoroughness becomes:
	\begin{equation}\label{eq:8}
		\sum_{i=1}^{i_{max}} c(i) = \frac{\Delta p}{K} \cdot \frac{i_{max}(i_{max}+1)}{2} = \frac{\Delta p^{2}}{K_{2}} \cdot \frac{i_{max}^{2}+i_{max}}{2}\in \mathcal{O}(i_{max}^2)
	\end{equation}
	Now it seems interesting to check if the quadratic nature of the payment is preserved by our brand new $c(i)$:
	\begin{equation*}
		\sum_{i=1}^{i_{max}} c(i) = \sum_{i=1}^{i_{max}} \Delta p(i) \frac{p(i)-p(0)}{K_{2}} = \frac{1}{K_{2}} \sum_{i=1}^{i_{max}} \Delta p(i) \cdot \big[ p(i)-p(0) \big]
	\end{equation*}
	as previously seen, the difference inside square brackets can be expressed as summation of marginal probabilities $\Delta p(i)$, so:
	\begin{equation}\label{eq:9}
		\sum_{i=1}^{i_{max}} c(i) =  \frac{1}{K_{2}} \sum_{i=1}^{i_{max}} \Delta p(i) \sum_{j=1}^{i} \Delta p(j) =  \frac{1}{K_{2}} \sum_{i=1}^{i_{max}} \sum_{j=1}^{i} \Delta p(i) \Delta p(j)
	\end{equation}
	We need to untie that double summation. Let’s study $i-j$ plane to understand what we are adding up:
	\begin{figure}[H]	
		\centering
		\includegraphics[height=0.33\textheight]{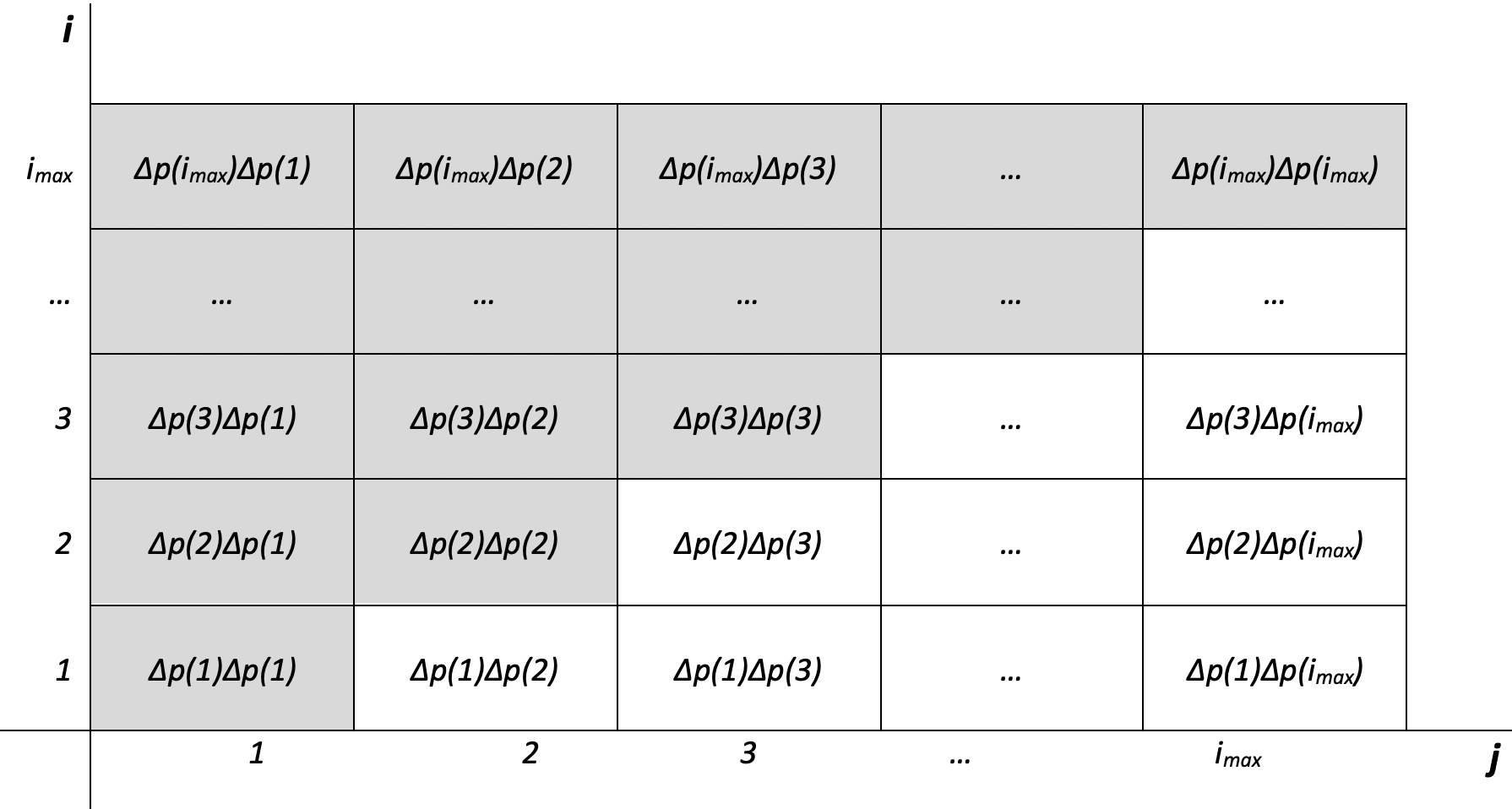}
		\caption{$i-j$ plane}
		\label{fig:ijplane}
	\end{figure}
	The double summation regards the gray cells; more, values of cells are symmetric with respect to the diagonal, so when we deal with summation of values:
	\begin{equation}\label{eq:10}
		\textrm{gray cells} = (\textrm{whole plane}+\textrm{diagonal}) / 2
	\end{equation}
	Summation of diagonal is given by: $\sum \Delta p(i)^{2}$
	
	To calculate the sum of whole plane we note that, for example, for the
	lowest row we have:
	\begin{equation*}
		\Delta p(1) \cdot \big[ \Delta p(1)+ \Delta p(2) + \dots + \Delta p(i_{max}) \big] = \Delta p(1) \sum_{i=1}^{i_{max}}p(i)
	\end{equation*}
	and likewise for all other rows as well. So when we add all rows to get the whole plane:
	\begin{equation*}
		\big[ \Delta p(1)+\Delta p(2)+\dots+\Delta p(i_{max}) \big] \sum_{i=1}^{i_{max}}p(i) = \sum_{i=1}^{i_{max}}p(i) \sum_{i=1}^{i_{max}}p(i) = \left[ \sum_{i=1}^{i_{max}}p(i) \right]^{2}
	\end{equation*}
	Putting together \ref{eq:9} and \ref{eq:10} we get:
	\begin{equation}\label{eq:11}
		\sum_{i=1}^{i_{max}} c(i) = \frac{\left[ \displaystyle\sum_{i=1}^{i_{max}}p(i) \right]^{2} +  \displaystyle\sum_{i=1}^{i_{max}}p(i)^{2}}{2 K_{2}}
	\end{equation}
	or:
	\begin{equation}\label{eq:12}
		\sum_{i=1}^{i_{max}} c(i) = \frac{A(i_{max})^{2} \cdot i_{max}^{2} + B(i_{max}) \cdot i_{max}}{2 K_{2}}
	\end{equation}
	where we have defined the average values in $[1,i_{max}]$:
	\begin{equation}\label{eq:13}
		\begin{aligned}
			A(i_{max}) &\triangleq \left< \Delta p(i) \right> \quad \leq 1\\
			B(i_{max}) &\triangleq \left< \Delta p(i)^{2} \right> \quad \leq 1
		\end{aligned}
	\end{equation}
	Not surprisingly, when $\Delta p(i) \equiv \Delta p$ we obtain $A \equiv \Delta p$ , $B \equiv \Delta p^{2}$ and \ref{eq:12} falls back exactly to \ref{eq:8}.
	
	In general case, using \ref{eq:13} inequalities in \ref{eq:12}, we obtain:
	\begin{equation*}
		\sum_{i=1}^{i_{max}} c(i) \leq \frac{i_{max}^{2} + i_{max}}{2 K_{2}}
	\end{equation*}
	Applying Wikipedia formal definition of Big-$\mathcal{O}$ notation \cite{BigO}:
	\begin{quote}
		\definecolor{block-gray}{gray}{0.95}
		\newtcolorbox{barobackcolor}{colback=block-gray, boxrule=0pt,boxsep=0pt,breakable}
		\begin{barobackcolor}
		That is, $f(x)=\mathcal{O}\big(g(x)\bigr)$ if there exists a positive real number $M$ and a real number $x_{0}$ such that\\
		\\
		$\quad \quad \left| f(x) \right| \leq Mg(x) \quad \textrm{for all } x \geq x_{0}$
		\end{barobackcolor}	
	\end{quote}
	we get:
	\begin{equation*}
		\sum_{i=1}^{i_{max}} c(i) \leq \frac{i_{max}^{2} + i_{max}}{2 K_{2}} \leq M \cdot i_{max}^{2} \quad \Longrightarrow \quad i_{max} \geq \frac{1}{2K_{2}M-1}
	\end{equation*}
	So we could still talk of quadratic payments in the sense that:
	\begin{equation*}
		\sum_{i=1}^{i_{max}} c(i) \in \mathcal{O}(i_{max}^2)
	\end{equation*}
	But remembering that 1 is just an upper-bound (a very high upper-bound probably) for both $A$ and $B$, above result is perhaps too coarse: let’s check with some graphs.
	These are four different probability functions $p(i)$ and their cost functions $c(i)$ for $1 \leq i \leq 1000$:
	\begin{figure}[H]
		\begin{subfigure}{0.5\linewidth}
			\includegraphics[width=\linewidth]{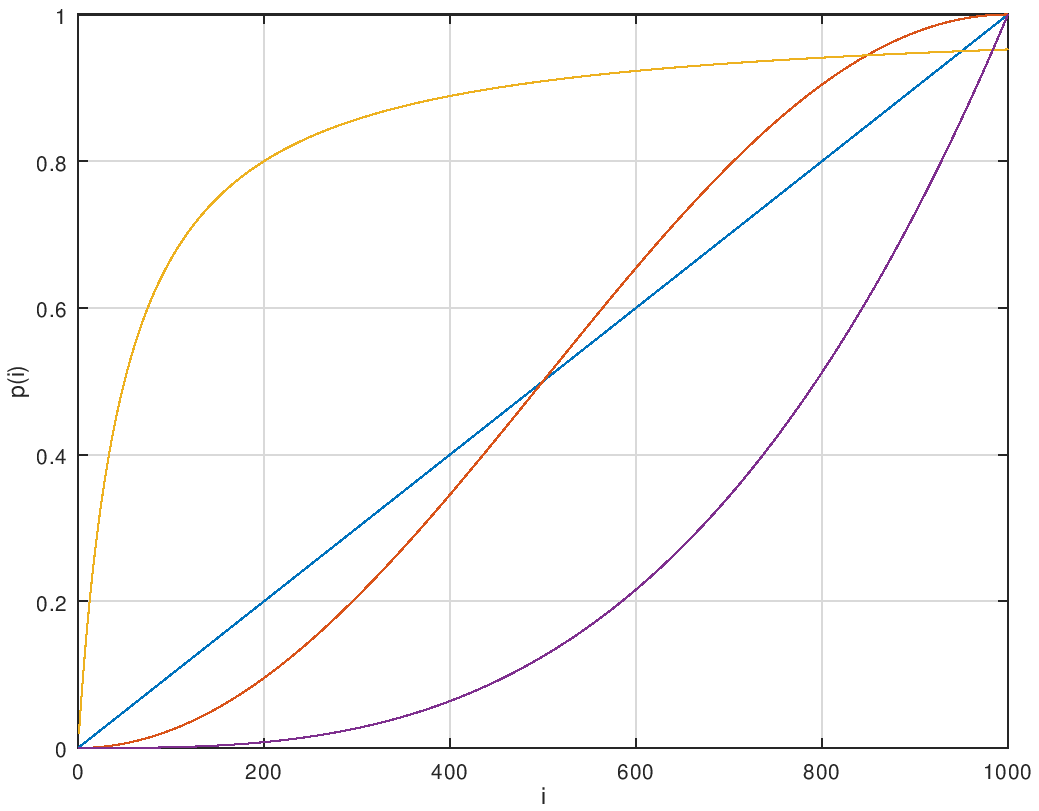}
		\end{subfigure}
		\begin{subfigure}{0.5\linewidth}
			\includegraphics[width=\linewidth]{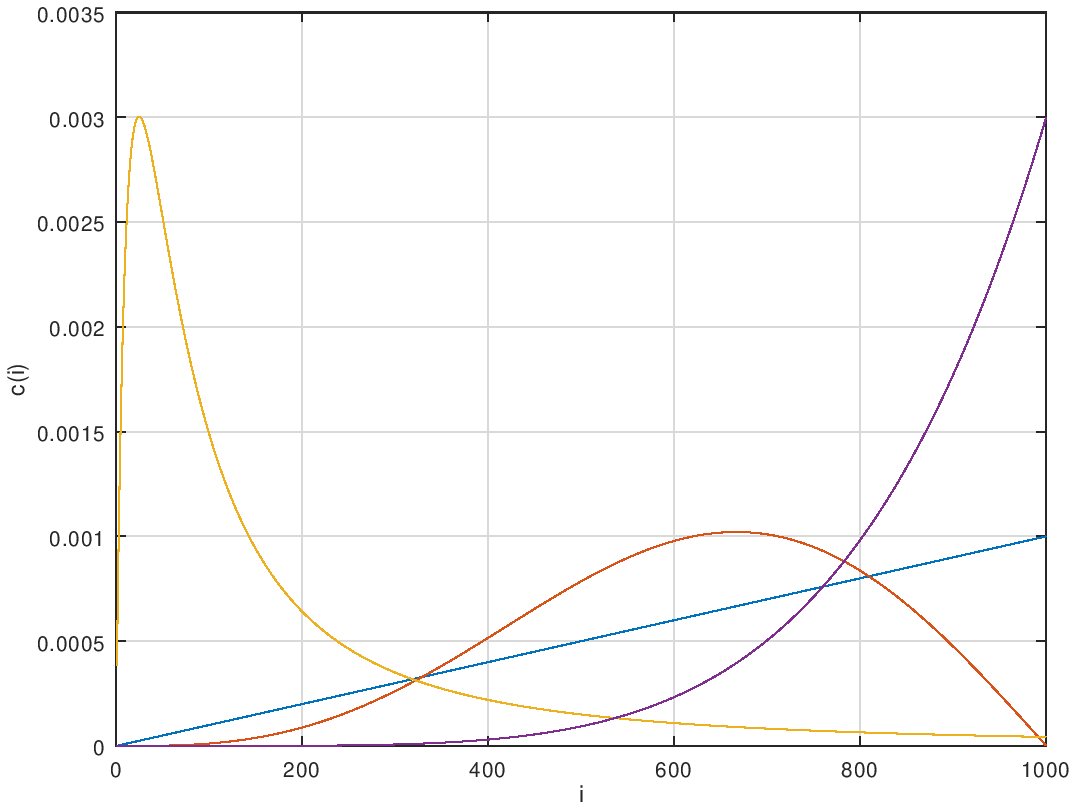}
		\end{subfigure}
		\caption{Miscellaneous probability functions \& related cost ones}
		\label{fig:misccp}
	\end{figure}
	Their total cost functions for a given $i_{max}$ is:
	\begin{figure}[H]	
		\centering
		\includegraphics[height=0.33\textheight]{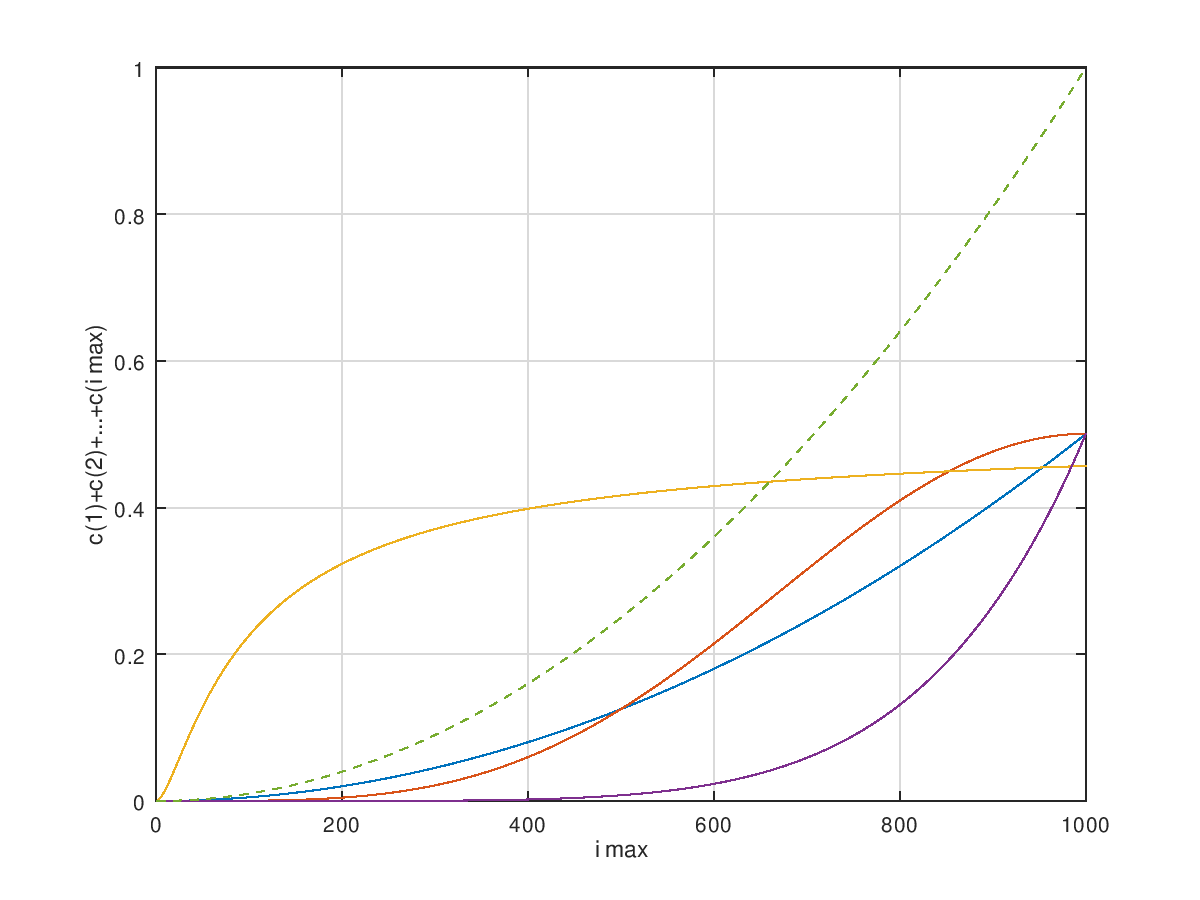}
		\caption{Miscellaneous total cost functions}
		\label{fig:misctotc}
	\end{figure}
	The dashed line is a reference $M \cdot i_{max}^{2}$ : as we have been suspecting, it’s definitely not a good estimator of total cost $\sum c(i)$ behavior for a generic probability $p(i)$.
	However it jumps to the eye that total cost functions aren’t too different from their own probability functions. This suggests that perhaps Big-$\mathcal{O}$ equivalence classes should be driven by $p(i)$ and not by $i$: let’s take \ref{eq:11} and rewrite it as:
	\begin{equation*}
		\sum_{i=1}^{i_{max}} c(i) = \frac{\left[ \displaystyle\sum_{i=1}^{i_{max}}p(i) \right]^{2} +  \displaystyle\sum_{i=1}^{i_{max}}p(i)^{2}}{2 K_{2}} = \frac{\big[ p(i_{max})-p(0)\big]^{2} +  \displaystyle\sum_{i=1}^{i_{max}}p(i)^{2}}{2 K_{2}}
	\end{equation*}
	Remembering that $\Delta p(i) \leq 1 \Longrightarrow \Delta p(i)^{2} \leq \Delta p(i)$ :
	\begin{equation*}
		\sum_{i=1}^{i_{max}} c(i) \leq \frac{\big[ p(i_{max})-p(0)\big]^{2} +  \displaystyle\sum_{i=1}^{i_{max}}p(i)}{2 K_{2}} = \frac{\big[ p(i_{max})-p(0)\big]^{2} + p(i_{max})-p(0)}{2 K_{2}}
	\end{equation*}
	From algebra:
	\begin{equation*}
		\sum_{i=1}^{i_{max}} c(i) \leq \frac{p(i_{max})^{2} + \big[1-2p(0)\big] p(i_{max}) +p(0)^{2} - p(0)}{2 K_{2}}
	\end{equation*}
	But, again, $p(0)^{2} \leq p(0)$ , so:
	\begin{equation*}
		\sum_{i=1}^{i_{max}} c(i) \leq \frac{p(i_{max})^{2} + \big[1-2p(0)\big] p(i_{max})}{2 K_{2}} \leq M \cdot p(i_{max})^{2}
	\end{equation*}
	If $[1-2p(0)] \leq 0$ the above inequality will be easily satisfied, for example, by $M=1/2K_{2} \quad \forall \ i_{max}$
	
	When instead [1-2p(0)] > 0 we will have:
	\begin{equation*}
		p(i_{max}) \geq \frac{1-2p(0)}{2K_{2}M-1} \quad \Longrightarrow \quad i_{max} \geq p^{-1}\left( \frac{1-2p(0)}{2K_{2}M-1} \right)
	\end{equation*}
	with, for the inverted probability argument, the constraint:
	\begin{align*}
		0 \leq \frac{1-2p(0)}{2K_{2}M-1} \leq 1 \quad &\Longrightarrow \quad
		\begin{cases}
			2K_{2}M-1 > 0 &\Longrightarrow M > 1/2K_{2}\\
			1-2p(0) \leq 2K_{2}M-1 &\Longrightarrow M \geq \displaystyle\frac{1-p(0)}{K_{2}}
		\end{cases}\\
		&\Longrightarrow \quad M \geq \displaystyle\frac{1-p(0)}{K_{2}}
	\end{align*}	
	So we can say:
	\begin{equation*}
		\sum_{i=1}^{i_{max}} c(i) \in \mathcal{O}\big(p(i_{max})^2\big)
	\end{equation*}
	which is a far better estimator than earlier one. So \enquote{legacy} quadratic payments (QP) actually seem a particular case of more general family which could perhaps be named \textit{quadratic success payments} (QSP) - because, remember, $p(i)$ is the probability of desired outcome given $i$ involvement (e.g. bought votes).
	
	\section{Recapping}
	Our long journey can be summarized by following table:
	\begin{table}[H]
		\begin{center}
			\begin{tabular}{l|l}
				\toprule
				generic $\Delta p(i)$ & $\Delta p(i) \equiv \Delta p$ \\
				\midrule
				& \\
				$i_{max}=\left\lfloor p^{-1}(K_{2}V+p(0)) \right\rfloor$ \qquad \qquad & $i_{max}=\left\lfloor KV \right\rfloor$ \\
				& \qquad \qquad \qquad \quad with $K=K_{2}/\Delta p$\\
				$c(i)=\Delta p(i) \displaystyle\frac{p(i)-p(0)}{K_{2}}$ \qquad \qquad & 
				$c(i) = \displaystyle\frac{\Delta p}{K} i$ \\
				& \\
				$\displaystyle\sum_{i=1}^{i_{max}} c(i) \in \mathcal{O}\big(p(i_{max})^2\big)$ \qquad \qquad & $\displaystyle\sum_{i=1}^{i_{max}} c(i) \in \mathcal{O}\big(i_{max}^2\big)$ \\
				& \\
				\bottomrule 
			\end{tabular}
			\caption{Recapping QSP \& QP}
			\label{tab:recapping}
		\end{center}
	\end{table}
	These constraints apply:
	\begin{enumerate}[label=\alph*)]
		\item $i$ and $i_{max}$	are whole numbers
		\item $p(i)$ is monotonically strictly increasing
		\item $p(i):\mathbb{R} \rightarrow \mathbb{R}$ (but also $p(i):\mathbb{N} \rightarrow \mathbb{R}$ extended by a polyline is ok)
		\item $K_{2}>0$
		\item $K_{2}<\big(1-p(0)\big)/V$
	\end{enumerate}

	Constraint a) is a way of saying that involvement of stakeholders has its own intrinsic granularity: e.g. you can buy 1 vote, 2 votes, $n$ votes but you cannot buy 1.5 votes. Note that it doesn’t rule out decimals altogether: there could be contexts in which involvement envisages fractions of unit, but those cases can fall back to whole numbers via rescaling. However continuity of real numbers isn’t contemplated;
	
	constraints b) and c) formalize obvious characteristics of a probability function linking desired outcome to involvement, and guarantee its invertibility;
	
	constraint d) is what makes influence proportional to perceived value $V$ (and being both positives, $K_{2}$ also has to);
	
	constraint e) tells us that exercisable infuence is a finite asset: from \ref{eq:6} and \ref{eq:7} we know that influence is strictly linked to product $K_{2}V$, so having the upper-bound of $K_{2}$ inversely proportional to $V$ means maximum influence is limited as well (to 1-p(0) not surprisingly, given our definition of influence related to marginal probability).\\
	And there’s also a consequence on market definition because it’s a two-way constraint between $K_{2}$ and $V$:
	\begin{itemize}
		\item $V$ is a per-stakeholder parameter, meaning that each player will have a different perception of desired outcome value;
		\item given $p(i)$, $K_{2}$ has an important role in defining cost function $c(i)$, which hasn’t to be per-stakeholder-defined: otherwise incentives mechanics would not succeed in inducing fair ratios between players’ influences, given their perception of value (if cost function changes, two stakeholders with the same V could have different $i_{max}$ and, consequently, different acquired influences). So we need a single $K_{2}$ for the entire market;
	\end{itemize}
	all of that seems to mean we should choose $K_{2}$ as small as we can to raise the bar of allowed $V$, however:
	\begin{itemize}
		\item a hugely motivated player we haven’t foreseen (aka one with a very high perceived value $V$, higher than our predictions) will always be able to fall out of our game rules;
		\item prices would increase ($K_{2}$ is at denominator of $c(i)$), meaning that stakeholders with small $V$s would be compressed towards small $i_{max}$ values: being whole numbers, the risk of too much compression is to flatten players with different perceived values to the same $i_{max}$.
	\end{itemize}
	Elaborating a bit more: when you choose $K_{2}$ , e) gives you an upper-bound for $V$, meaning that a stakeholder with that maximum $V$ will be incentivized to acquire all available influence ($1-p(0)$): so we should choose a $K_{2}$ corresponding to the sum of all $V$s of all players (if predictable) plus some \enquote{guard value} to take into account unexpected stakeholders appearance. How small $K_{2}$ (or big the \enquote{guard value}) will be is a trade-off needing to balance:
	\begin{itemize}
		\item how much protection is needed against a player with a very big $V$ coming earlier than others and hoarding influence;
		\item the need to keep prices low enough to have a good spread of $i_{max}$ values for actual players.
	\end{itemize}
	So $K_{2}$ also affects which difference between two values of $V$ is big enough to guarantee two different values of $i_{max}$, i.e. it chooses the granularity of $V$: because no rational player will pay more than the minimum needed for a certain level of influence.
	\begin{table}[H]
		\begin{center}
			\begin{tabular}{c|c}
				\toprule
				big $K_{2}$ & small $K_{2}$ \\
				\midrule
				& \\
				More spread of $i_{max}$/influence (finer-grained $V$) & Bigger $V$ allowed \\
				& \\
				\bottomrule 
			\end{tabular}
			\caption{$K_{2}$ vs $V$ trade-off}
			\label{tab:tradeoff}
		\end{center}
	\end{table}
	In optimal case it will be:
	\begin{equation*}
		\min_{m \neq n}\Big| \left\lfloor p^{-1}(K_{2}V_{m}+p(0)) \right\rfloor - \left\lfloor p^{-1}(K_{2}V_{n}+p(0)) \right\rfloor \Big| = 1
	\end{equation*}
	Of course we have always been assuming that $p(i)$ saturates to 1 for an $i$ big enough to have \enquote{space} permitting fair involvement of all interested players: it doesn’t seem a limiting condition because if there isn’t enough \enquote{space to act} then there isn’t a market either.
	
	\section{Not only referendums}
	During last thoughts about constraints I have mainly used general terms: e.g. market and stakeholders instead of ballot and voters: that’s because formulas scope seems wider than the initial referendum context.
	
	That’s why -for example- previous words about maximum allowed sum of $V$s seem to suggest stakeholders could be in mutual competition buying as much influence as possible, so to be worried about voters with high $V$s: it’s a general possibility emerging from the framework, but it isn’t always that case, and probably it isn’t when we are modeling voters hoping for the same event (\enquote{yes} winning): granularity of $V$ seems much more important to exploit each voter commitment fairly.
	
	So, reconsidering what we are dealing with from a more general point of view, we have:
	\begin{itemize}
		\item a stochastic process $S$ whose execution could lead or not to $D$ outcome;
		\item an $i$ metric of involvement in favor of $D$;
		\item a generic $p(i)$ probability of $D$ outcome;
		\item a $V$ value assigned to $D$ outcome by a stakeholder;
		\item a pricing of $i$ with a cost function $c(i)$ incentivizing the stakeholder to influence the outcome in favor of $D$ (by means of $i$ buying) in a proportional-to-$V$ way: greater involvement would be anti-economic.
	\end{itemize}
	The whole picture seems to suggest many applications... but that’s enough for now! Thanks for reading!
\end{document}